\documentclass[prb,showpacs,twocolumn,final,superscriptaddress]{revtex4}

\usepackage{amsfonts,amsmath,amssymb,ifpdf,epsfig,graphicx,color}

\newcommand{\beq}{\begin{equation}}
\newcommand{\eeq}{\end{equation}}
\newcommand{\ua}{\uparrow}
\newcommand{\da}{\downarrow}
\newcommand{\Eq}[1]{Eq.~(\ref{#1})}

\newcommand{\M}[1]{\mathcal{M}^{(#1)}}
\newcommand{\tave}{t_{\text{\tiny av}}}
\newcommand{\trel}{t_{\text{\tiny rel}}}

\begin{document}

\title{Non-equilibrium steady state in a periodically driven Kondo model}

\author{M. Heyl}

\author{S. Kehrein}

\affiliation{Physics Department, Arnold Sommerfeld Center for Theoretical Physics and Center for NanoScience, \\
Ludwig-Maximilians-Universit\"at, Theresienstrasse 37, 80333 Munich, Germany}

\begin{abstract}
We investigate the Kondo model with time-dependent couplings that are periodically
switched on and off. On the Toulouse line we derive exact analytical results for the spin dynamics
in the steady state that builds up after an infinite number of switching periods. Remarkably, the universal
long-time behavior of the spin-spin correlation function remains completely unaffected by the driving. 
In the limit of slow driving the dynamics becomes equivalent to that of a single interaction quench.
In the limit of fast driving it is shown that the steady state cannot be described by some effective
equilibrium Hamiltonian due to the observation that an incautious implementation of the Trotter
formula is not correct. As a consequence, the steady state in the limit of fast switching serves as an example for the emergence of new quantum states not accessible in equilibrium.
\end{abstract}

\date{\today}

\pacs{72.15.Qm,85.35.Be,73.50.Mx}

\maketitle

\section{Introduction}

Recent progress in experiments stimulated the interest in non-equilibrium phenomena of interacting many-particle systems. Cold atoms trapped in optical lattices offer the possibility of studying the time evolution of quantum many-body systems with time-dependent system parameters.\cite{Optical_lattices} Due to the excellent isolation from the environment the non-equilibrium dynamics of these systems are accessible with negligible decoherence over long times.

Nanostructures such as quantum dots provide the framework to examine experimentally the non-equilibrium dynamics in quantum impurity models. Most importantly for the present work, quantum dots can act as magnetic impurities displaying Kondo physics.\cite{Exp_Kondo} In contrast to impurities in a bulk sample, unscreened electrical or magnetic fields can be applied directly such that the local system parameters can be varied in time by choosing appropriate time-dependent fields.\cite{Nordl_ac}

The possibility to experimentally study the properties of interacting many-body systems out of equilibrium motivated numerous analytical and numerical theoretical treatments. Most of the activities have been concentrating on interaction quenches in various model systems.\cite{Lob_dyn,Lob_eff,quenches,Nordl_Kondo_develop} For periodically driven interacting many-body systems, however, less results are known. Recently, periodic time-dependent Falicov-Kimball models have been investigated in the limit of infinite dimensions by using dynamical mean-field theory.\cite{FalKim} Considerable activity in the field of time-dependent quantum impurity models led to a number of works on periodically driven Anderson impurity\cite{driven_Anderson,Nordl_ac} and Kondo models\cite{Kam_prl,Kam_prb,Goldin,Schiller_Hershfield}.

As the Kondo model is the paradigm model for strongly correlated systems, it is of particular interest in the field of non-equilibrium phenomena. In equilibrium, the Kondo effect emerges from the interaction of a localized spin degree of freedom with a bath of surrounding electrons. At sufficiently low temperatures, this bath of itinerant electrons develops a localized spin polarization cloud in the vicinity of the local spin, the so-called Kondo cloud, providing a mechanism to screen the local magnetic moment. In the zero temperature limit, the screening becomes dominant leading to the emergence of a bound state. The surrounding spin polarization cloud is tied to the local spin establishing the so-called Kondo singlet with an associated binding energy $T_K$, the Kondo temperature. The Kondo effect manifests itself most prominently in the Kondo resonance, a sharp peak in the local density of states that is pinned exactly at the Fermi energy. As the Kondo effect is a coherent many-body phenomenon, the question arises how it is affected in a non-equilibrium setting.

Due to the complexity of many-body systems out of equilibrium, it is instructive to investigate those cases where exact nonperturbative solutions are accessible. In this context, it is of particular interest that the Kondo model as a paradigm model for strongly correlated systems exhibits a special line in parameter space, the Toulouse limit, where it becomes exactly solvable.\cite{Toulouse} The Toulouse limit displays many generic and universal properties of the strong coupling limit of the Kondo model in equilibrium as well as for interaction quenches~\cite{Lob_dyn}. The local spin dynamics, for example, that is also investigated in this work, is well described whereas other universal quantities such as the Wilson ratio explicitly depend on the anisotropy.

The exact solvability of the Kondo model in the Toulouse limit is used in this work to investigate nonperturbatively a steady state that is generated by periodically switching on and off the interaction at zero temperature. This steady state is characterized by analyzing exact analytical results for the local dynamical quantities, that is the magnetization of the impurity spin, the spin-spin correlation function and the dynamical spin susceptibility.

A system that is driven by an external force approaches a steady state if the amount of energy that is provided to the system does not lead to an overheating, as it may happen for systems with an unbounded spectrum. Since these steady states emerge from a non-equilibrium setting, equilibrium thermodynamics is not applicable for their description. As a consequence, these states display new properties that are not accessible by exciting the system thermally. Their characterization, however, poses a new challenge. Recently, there have been attempts to assign effective thermodynamic quantities such as effective temperatures to describe the properties of systems in a nonequilibrium setup.\cite{Lob_eff,Nordl_Kondo_develop,Nordl_ac,Mitra} As will be shown in  this paper, a characterization of the present steady state in terms of an effective temperature is not possible. The excitations that are created by the periodic driving are fundamentally different from those induced by temperature. A finite temperature smears the Fermi surface whereas the periodic driving leads to an excitation spectrum of discrete character with excitations of multiples of the driving frequency.

Another question arising in the context of driven systems is whether the universality in equilibrium systems such as the Kondo model extends to the non-equilibrium case, whether new universal properties emerge and whether equilibrium quantities remain meaningful. Kaminski and coworkers\cite{Kam_prl,Kam_prb}, for example, proposed a universal description for the conductance through a Kondo impurity for a periodically driven Kondo model. In the present setting, the spin-spin correlation function displays a universal long-time behavior that is completely independent of the driving. This universality originates from the fact that the low-energy excitations in the immediate vicinity of the Fermi level that are relevant for the long-time behavior are unaffected by the periodic driving due to the discrete character of the excitation spectrum.

This paper is organized as follows. In Sec.~II, the model Hamiltonian, a time-dependent Kondo model in the Toulouse limit, is introduced and mapped onto a quadratic effective Hamiltonian. The method used to determine the time evolution in the periodic driving setup is presented in Sec.~III. The result for the magnetization of the impurity spin is shown in Sec.~IV. Sec.~V is devoted to a detailed analysis of the spin-spin correlation function and the results for the dynamical spin susceptibility are presented in Sec.~VI. 

\begin{figure}
	\includegraphics[width=0.95\linewidth]{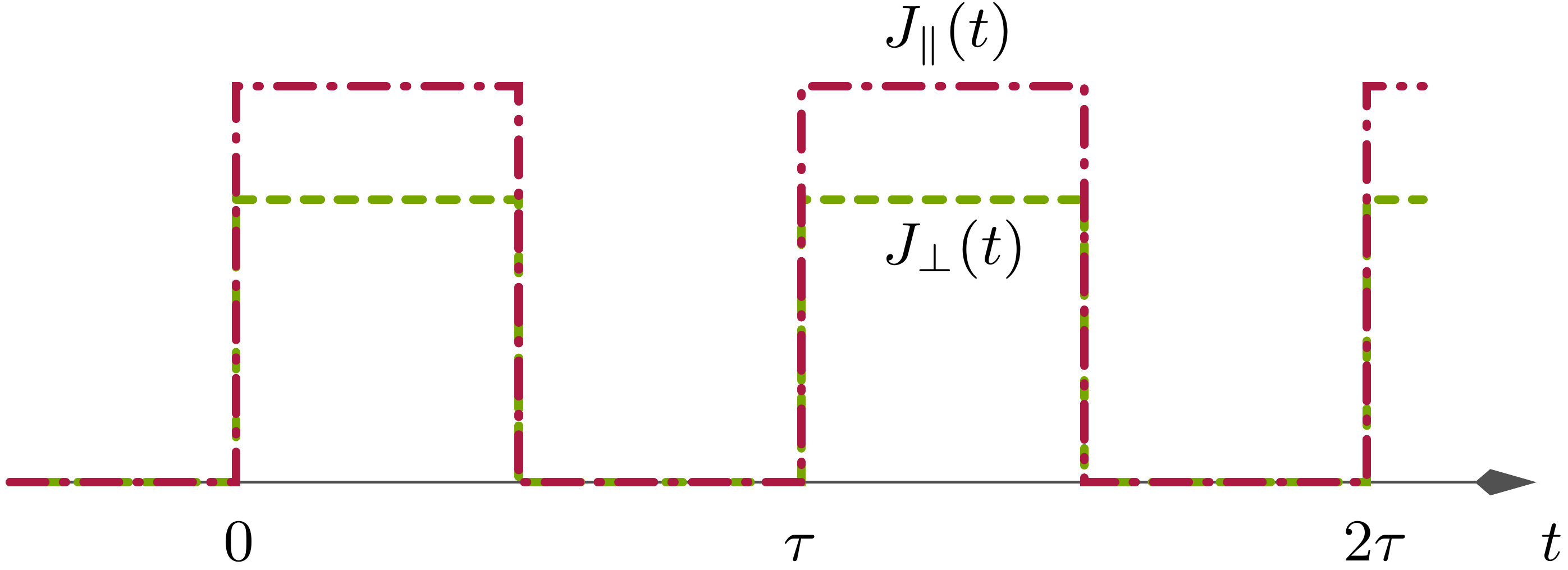}
\caption{Time dependence of the perpendicular $J_\perp(t)$ and parallel coupling $J_\parallel(t)$ in the anisotropic Kondo Hamiltonian. For times $t<0$ all couplings are zero whereas for $t>0$ they are switched on and off periodically with period $\tau$.}
\label{pic_switching}
\end{figure}

\section{Periodic time-dependent Kondo model}

Consider a local spin whose exchange interaction with the surrounding electrons is switched on and off periodically, which generates a Kondo Hamiltonian with time-dependent couplings. For convenience we allow for an anisotropy in the exchange interaction leading to different couplings in $z$ direction, $J_z=J_\parallel$, and in the $xy$-plane, $J_x=J_y=J_\perp$:
\beq
	H=\sum_{k \alpha} k \mbox{:} c_{k\alpha}^\dag c_{k\alpha} \mbox{:} + \sum_{i} \frac{J_i(t)}{2} \sum_{\alpha,\alpha'} \, \mbox{:} \Psi_{\alpha}^\dag(0) \sigma_i^{\alpha,\alpha'} S_i \Psi_{\alpha'}(0)  \mbox{:}.
\label{AIKM}
\eeq
The operator $c_{k\alpha}^\dag$ creates an electron with wave vector $k$ and spin $\alpha=\ua,\da$ in the reservoir. The colons $:\ldots:$ denote normal ordering relative to the Fermi sea. The local spin operator $\vec{S}$ with components $S_i,i=x,y,z,$ is coupled to the local spin density of the conduction-band electrons whose components are determined by the Pauli matrices $\sigma_i$. The electron's dispersion relation has been linearized around the Fermi level and energies are measured in units of $v_F$ relative to the Fermi energy, i.e., $v_F=1$ and $\varepsilon_F=0$. As the local scatterer is assumed to be pointlike, only $s$-wave scattering occurs rendering the problem to be effectively one-dimensional.\cite{Leggett}

For negative times, the system is prepared in one of the ground states $|\Psi_0\rangle$ of the noninteracting problem that are product states of the Fermi sea for the conduction-band electrons and a wave function for the local spin. As depicted in Fig.~\ref{pic_switching}, at time $t=0$, the periodic driving process starts by switching on the interaction. After half of the period $\tau$, $t=\tau/2$, the interaction is switched off until $t=\tau$. Afterwards, this procedure is continued, until after an infinite number of periods a steady state builds up, a state in which all real-time correlation functions are invariant under a discrete time shift of one period in all their time arguments. The energy scale associated with the periodic switching is the driving frequency $\Omega=2\pi/\tau$.

Due to the time-dependence of the Hamiltonian, energy is not a conserved quantity. Moreover, each quench that is performed excites the system such that one can expect that after an infinite number of switchings, in the steady state, an infinite amount of energy is pumped into the system. This is indeed the case in the present setup. Therefore, a dissipation mechanism is needed that prevents the system from overheating. As has been shown by Doyon and Andrei\cite{Doyon}, the conduction band in the Kondo model itself can serve as a bath if and only if it is taken as infinitely large. Therefore, a definite order of taking limits has to be prescribed, namely, the thermodynamic limit has to be taken before the limit of long times:
\beq
	\lim_{t\to \infty} \lim_{L\to \infty}.
\eeq
In this way, as expected the conduction band is prevented from overheating induced by a single impurity.

One possible way of experimentally realizing the periodic switch on and off of the Kondo interaction in a quantum dot is the following: consider a quantum dot with a local single-particle level at energy $\varepsilon_d$ and a large on-site interaction $U$ in the Kondo regime where $\varepsilon_d \ll \varepsilon_F, \varepsilon_d+U \gg \varepsilon_F$ and $|\varepsilon_d-\varepsilon_F|,|\varepsilon_d+U-\varepsilon_F| \gg \Gamma,T$. Here, $\Gamma$ denotes the broadening of the local level $\varepsilon_d$ in the quantum dot and $T$ the temperature. Via the Schrieffer-Wolff transformation, the corresponding Anderson impurity model can be mapped onto a Kondo model with an exchange coupling $J\propto\Gamma \left[ \frac{1}{\varepsilon_F-\varepsilon_d-U} + \frac{1}{\varepsilon_F-\varepsilon_d}\right]$.\cite{Schrieffer_Wolff} Following a suggestion by Nordlander and coworkers\cite{Nordl_Kondo_develop}, consider the case where the local single-particle level $\varepsilon_d$ alternates between two different values $\varepsilon_{d1}$ and $\varepsilon_{d2}$ with $|\varepsilon_{d2}-\varepsilon_F| \gg |\varepsilon_{d1}-\varepsilon_F|$ where for each $\varepsilon_{d1}$ and $\varepsilon_{d2}$ the quantum dot is assumed to be in the Kondo regime. Then, the Kondo exchange coupling $J_2$ corresponding to $\varepsilon_{d2}$ is much smaller than $J_1$. The associated Kondo temperature $k_B T_{K2}=D \sqrt{\rho_0 J_2}e^{-1/(J_2 \rho_0)}$, $\rho_0$ is the density of state at the Fermi level and $D$ a high-energy cutoff, vanishes exponentially such that $J_2$ can be set equal to zero.  As a result, the corresponding Kondo model becomes time-dependent with an exchange interaction $J_1$ that is switched on and off periodically. This can be shown rigorously by performing a time-dependent Schrieffer-Wolff transformation.\cite{Goldin,Kam_prl,Kam_prb}

The periodic driving in a quantum dot sets an upper bound on the driving frequency $\Omega$. As the Kondo Hamiltonian requires strict single occupancy, the driving has to be small enough not to induce charge fluctuations on the dot caused by hopping processes between the central region and the conduction band by absorbing or emitting quanta of the driving frequency, i.e., $\Omega \ll |\varepsilon_d|,\varepsilon_d+U$.\cite{Kam_prl,Kam_prb,Goldin}

At zero temperature in equilibrium, the screening of the local magnetic moment by the conduction-band electrons becomes dominant leading to the emergence of a bound state called the Kondo singlet. Many universal features of the equilibrium Kondo model in this strong coupling limit are well described by the anisotropic Kondo Hamiltonian in the Toulouse limit that corresponds to a special line in parameter space of the Hamiltonian in \Eq{AIKM} where $J_\parallel=2-\sqrt{2}$. For this value of the parallel coupling, the Hamiltonian can be mapped onto an exactly solvable quadratic noninteracting resonant-level model using bosonization and refermionization.\cite{Leggett} Recently, it was shown by Lobaskin and Kehrein~\cite{Lob_dyn} that these methods can also be adopted to an exact solution of an interaction quench scenario. The only difference is an additional potential scattering term in the effective Hamiltonian.

The bosonization technique establishes a bosonic representation of fermionic fields $\Psi_\alpha(x)$ in one dimension called the bosonization identity, see Ref.~\cite{Delft_Bos} for a recent review. In the thermodynamic limit, the bosonization identity reduces to:
\beq
	\Psi_\alpha(x)=\frac{1}{\sqrt{a}} F_\alpha e^{-i\phi_\alpha(x)}, \:\: \alpha=\ua,\da,
\eeq
where the bosonic field $\phi_\alpha(x)=-\sum_{q>0}[e^{-iqx}b_{q\alpha}+e^{iqx}b_{q\alpha}^\dag]e^{-aq/2}/\sqrt{n_q}$ is related to the fermionic densities $\rho_\alpha(x)=\mbox{:}\Psi_\alpha^\dag(x)\Psi_\alpha(x)\mbox{:}=\partial_x \phi_\alpha(x)$ and $a^{-1}$ is an ultraviolet cutoff. The bosonic operators $b_{q\alpha}^\dag=i/\sqrt{n_q} \sum_{k}c_{k+q \alpha}^\dag c_{k\alpha}$ create a superposition of particle-hole pairs with momentum transfer $q=2\pi n_q/L>0$. Here, $L$ denotes the system size. The Klein factor $F_\alpha$ accounts for the annihilation of one electron as this cannot be achieved by the bosonic field $\phi_\alpha(x)$. 

By performing a sequence of unitary transformations, the Kondo Hamiltonian in the Toulouse limit can be simplified tremendously. First, the spin and charge $(s,c)$ degrees of freedom are separated by defining the bosonic fields $\phi_s(x)=[\phi_\ua(x)-\phi_\da(x)]/\sqrt{2}$ and $\phi_c(x)=[\phi_\ua(x)+\phi_\da(x)]/\sqrt{2}$. The charge sector of the anisotropic Kondo Hamiltonian is decoupled from the local spin and reduces to a collection of uncoupled harmonic oscillators. Therefore it will be omitted from now on. In the first half period, the interaction part in the spin sector $ J_\parallel/\sqrt{2} \partial_x \phi_s(x) S_z+J_\perp/(2a)[F_\ua^\dag F_\da e^{i\sqrt{2} \phi_s(0)} S_- +\text{h.c.} ] $ is modified by an Emery-Kivelson transformation $U=e^{i\gamma \phi_s(x)S_z}$, $\gamma=\sqrt{2}-1$, to $J_\perp/(2a)[F_\ua^\dag F_\da e^{i \phi_s(0)} S_- +\text{h.c.} ] $ in the Toulouse limit where $J_\parallel=2-\sqrt{2}$. For the second half period, the Emery-Kivelson transformation generates a scattering term $\propto \langle S_z(t) \rangle \partial_x \phi_s(0) $, whose strength depends on the instantaneous magnetization of the impurity spin $\langle S_z(t) \rangle$.\cite{Lob_dyn} The exponentials appearing in the transformed interaction part can be refermionized by introducing new spinless fermionic fields $\Psi(x)=a^{-1/2}F_s e^{-i\phi_s(x)}$ using the inverse of the bosonization identity where $F_s=F_\da ^\dag F_\ua$.\cite{Zarand_Delft} One may think of the $\Psi^\dag(x)$ fields as creating spin excitations at point $x$ in the reservoir. Moreover, another unitary transformation $U_2=e^{i\pi\mathcal{N}_s S_z}$ has to be imposed in order to arrive at a completely refermionized Hamiltonian and to ensure correct anticommutation relations for all operators.\cite{Zarand_Delft} Here, $\mathcal{N}_s=\frac{1}{2}[N_\ua-N_\da]$ measures the total spin polarization of the conduction-band electrons. By defining the operator $d=e^{-i\pi[\mathcal{N}_s-S_z]}S_-$ and its Hermitian conjugate $d^\dag$ as well as by performing a mode expansion for the new fermionic fields $c_k=(2\pi L)^{-1/2} \int dx \Psi(x) e^{ikx}$, one arrives at the following Hamiltonian:
\begin{eqnarray}
\label{eff_Hamiltonian}
	H=&& \sum_k k  \, \mbox{:} \, c_k^\dag  c_k  \, \mbox{:} \, + g(t) \langle S_z(t) \rangle \sum_{kk'}  \, \mbox{:} \, c_k^\dag c_{k'}  \, \mbox{:} \nonumber \\
	&&+ V(t) \sum_k \left[ c_k^\dag d + d^\dag c_k \right]
\end{eqnarray}
where $g(t)=g\theta(-\sin(\Omega t))$, $V(t)=V \theta(\sin(\Omega t))$, $V=J_\perp \sqrt{\pi/2aL}$, $g=(1-\sqrt{2})\pi/L$ for $t>0$, and $g(t)=g$, $V(t)=0$ for $t<0$. For times $N\tau<t<N\tau+\tau/2$ the Hamiltonian is a resonant-level model. For times $N\tau+\tau/2<t<N\tau+\tau$ the dynamics are governed by a potential scattering Hamiltonian and the local $d$ operators do not evolve in time such that $\langle S_z(t) \rangle=\langle S_z(N\tau+\tau/2)\rangle$. As it will turn out in the following analysis, the intermediate time evolution with the potential scattering Hamiltonian has no influence on the local spin dynamics such as the magnetization $\langle S_z(t)\rangle$ at all. Therefore, it is not necessary to solve the dynamics in a self-consistent way.

The Kondo scale can be connected to the parameters of the resonant-level model via the impurity contribution to the Sommerfeld coefficient in the specific heat:\cite{Lob_dyn} $C_{\text{\tiny imp}}=\gamma_{\text{\tiny imp}}T$ where $\gamma_{\text{\tiny imp}}=w \pi^2/3T_K$ and $w=0.4128$ is the Wilson number. In this way, the Kondo temperature $T_K$ is determined by $T_K=\pi w \Delta$ where $\Delta=V^2 L/2$ is the hybridization function.

The functional dependence between the spinless fermions $c_k$ and the conduction-band electrons $c_{k\alpha}$ is highly nonlinear and nontrivial. The local spin observable $S_z$, however, commutes with all unitary transformations and can be connected to operators of the effective Hamiltonian in a simple way:
\beq
\label{S_z_dd}
	S_z=d^\dag d-\frac{1}{2}.
\eeq
This relation allows to analytically calculate correlation functions that involve the $S_z$ observable such as the magnetization of the impurity spin $P(t)=\langle S_z(t) \rangle$, the spin-spin correlation function $\langle S_z(t) S_z(t') \rangle$ and the dynamical spin susceptibility $\chi''(t,\varepsilon)$, as it will be done in this work for the periodic driving setup.

\section{Time evolution}

\begin{figure}
\centering
	\includegraphics[width=\linewidth]{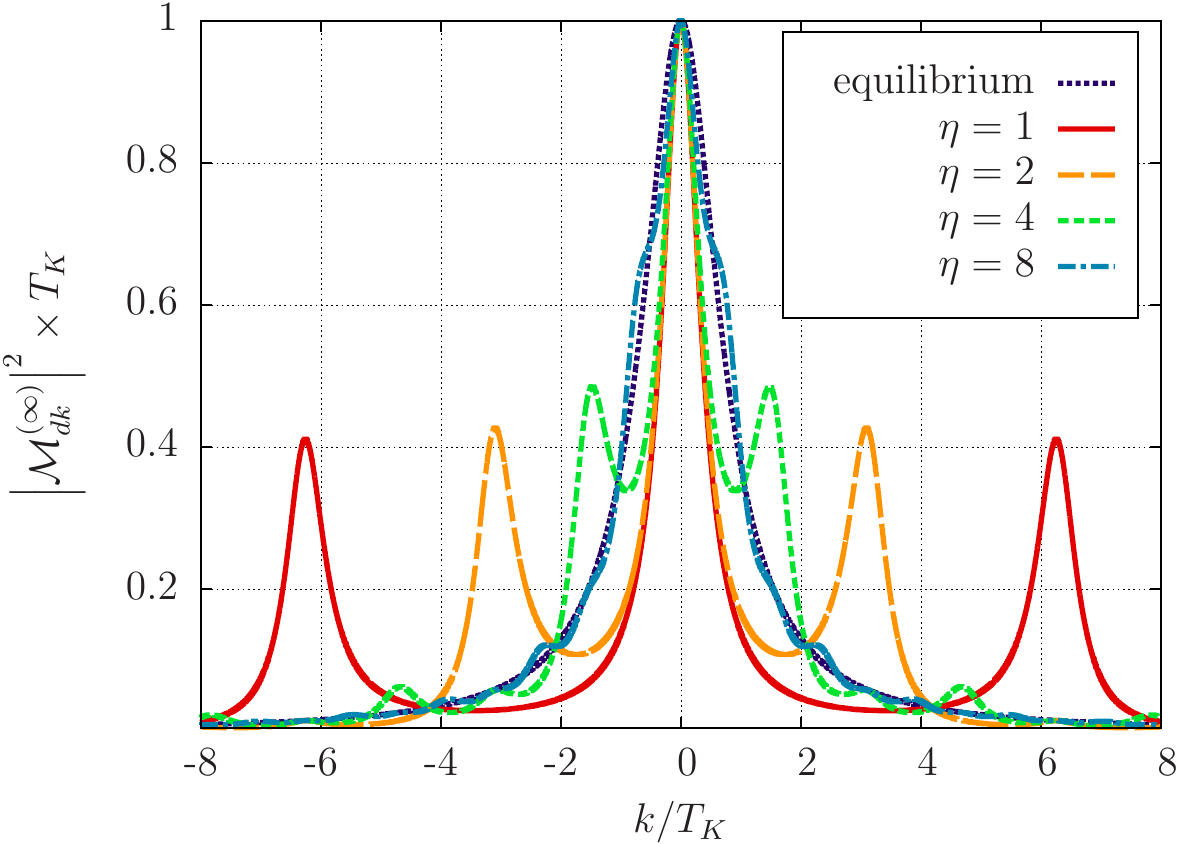}
\caption{Probability $|\mathcal{M}_{dk}^{(\infty)}|^2$ for a local $d$-fermion to decay into a bath state $k$ after an infinite number of periods. As a reference, the equilibrium curve for a resonant-level model with the same Kondo temperature $T_K$ is included. The parameter $\eta=\tau T_K=\tau/t_K$ compares the speed of switching $\tau$ with the internal time scale $t_K=1/T_K$.}
\label{pic_MkdN}
\end{figure}

As the Hamiltonian in \Eq{eff_Hamiltonian} is quadratic, the time evolution of the single-particle operators $c_k$ and $d$ is entirely determined by the Green's functions $G_{ll'}(t)=\theta(t) \langle \{ c_l(t) ,c_{l'}^\dag \}\rangle$:
\beq
	c_l(t)=\sum_{l'} G_{ll'}(t) c_{l'}, \: \: l,l'=k,d
\eeq
with a unitary matrix $G$. Despite the complexity of time evolution for time-dependent Hamiltonians, the periodic driving as it is considered in this work involves two time slices, during which the Hamiltonian is constant. For each half period, the dynamics is determined by the Green's functions for a potential scattering $\mathcal{G}^{(n)}$ and a resonant-level model $\mathcal{G}$. The Green's function for the potential scattering Hamiltonian varies for different periods, as the strength of the scatterer depends on the instantaneous magnetization of the impurity spin. Therefore, the Green's functions are labeled by an additional superscript where $n$ stands for the number of the period after starting the periodic driving. The time evolution over one period transforms the single-particle operators in the following way:
\beq
	c_l(\tau)=\sum_{l'} \mathcal{M}_{ll'}(n) c_{l'},\:\:\: \mathcal{M}_{ll'}(n)=\left[\mathcal{G}\left(\frac{\tau}{2}\right) \mathcal{G}^{(n)}\left(\frac{\tau}{2}\right) \right]_{ll'},
\eeq
defining the unitary matrices $\mathcal{M}(n)$ that are obtained by matrix multiplication. Note that the order of time evolution in the Heisenberg picture is opposite to the order of time evolution in the Schr\"odinger picture: the operators are first evolved according to the potential scattering Hamiltonian although it acts in the second half period. In this formulation, the problem of long-time evolution reduces to a matrix multiplication problem since the long-time evolution is completely determined by evolving the single-particle operators over multiple periods:
\beq
	c_l(N\tau)=\sum_{l'} \M{N}_{ll'} c_{l'},\:\:\: \M{N}_{ll'}=\left[\mathcal{M}(1)\dots \mathcal{M}(N)\right]_{ll'}.
\label{def_matrix_M_N}
\eeq
Finding the matrix elements of an arbitrary multiplication of some matrices can be a difficult task. The matrices $\mathcal{M}(n)$, however, display nice mathematical properties such that an analytical calculation can be carried out as is shown in Appendix~\ref{App_A}. The goal is to derive solvable recursion formulas by partially performing summations over intermediate indices. This yields the following relations:
\begin{widetext}
\begin{eqnarray}
\label{M_N}
	\M{N}_{dd}=e^{-N \Delta \tau/2}, \quad
	\M{N}_{dk}=\mathcal{M}_{dk} \frac{e^{-iNk\tau}-e^{-N\Delta \tau/2}}{e^{-ik\tau}-e^{-\Delta\tau/2}} ,\quad
	\M{N}_{kd}=\mathcal{M}_{kd} \frac{e^{-iNk\tau}-e^{-N\Delta \tau/2}}{e^{-ik\tau}-e^{-\Delta\tau/2}}, \nonumber \\
	\M{N}_{kk'}=\delta_{kk'}e^{-iNk\tau}+\mathcal{L}_{kk'} \frac{e^{-iNk\tau}-e^{-iNk'\tau}}{e^{-ik\tau}-e^{-ik'\tau}}
	+ \mathcal{M}_{kd}\mathcal{M}_{dk} \left[ \frac{e^{-iNk\tau}}{[e^{-ik\tau}-e^{-ik'\tau}][e^{-ik\tau}-e^{-\Delta\tau/2}]} + \right. \nonumber \\\
	+\left. \frac{e^{-N\Delta\tau/2}}{[e^{-ik\tau}-e^{-\Delta \tau/2}][e^{-ik'\tau}-e^{-\Delta\tau/2}]}+\frac{e^{-iNk'\tau}}{[e^{-ik'\tau}-e^{-ik \tau}][e^{-ik'\tau}-e^{-\Delta\tau/2}]} \right]+\mathcal{K}^{(N)}_{kk'}.
\end{eqnarray}
\end{widetext}
The precise definition of the functions appearing in these relations can be found in Appendix~\ref{App_A}. The matrix elements $\M{N}_{ll'}$ can be interpreted as the probability amplitudes for a fermion in a single-particle state $|l\rangle=c_l^\dag|\rangle$ to transform into $|l'\rangle$ after $N$ periods $\tau$. Here, $|\rangle$ denotes the true vacuum without any fermion. In Fig.~\ref{pic_MkdN}, plots for $|\M{\infty}_{dk}|^2$ are shown, that is the probability for a $d$ fermion to decay into a $k$ fermion after an infinite number of periods, i.e., as the steady state has developed. For large periods, the probability distribution approaches its equilibrium shape of a resonant-level model whereas for decreasing the period $\tau$ side-peaks appear located at odd multiples of the driving frequency $\Omega$. This corresponds to hopping processes under the absorption or emission of an odd number of quanta $\Omega$. Regarding the $d^\dag$ and $d$ operators as raising or lowering the local spin and $c_k^\dag$ as creating a spin excitation of energy $k$ in the fermionic reservoir, one can deduce from Fig.~\ref{pic_MkdN} that the elementary excitations caused by the periodic driving are those where the local spin is flipped by simultaneously creating spin excitations of energy $n\Omega$, $n$ odd. A Fourier series expansion of the $\theta(\sin(\Omega t))$-function in the interval $[0,\tau]$ reveals why mostly the odd frequencies contribute:
\beq
	\theta\left[\sin(\Omega t)\right]=-\sum_{n \in 2\mathbb{Z}+1} \cos(n \Omega t) .
\label{odd_freq}
\eeq
Therefore, the switch on and off driving can be thought of as a non-monochromatic driving including all odd multiples of the driving frequency.

As an aside, one can deduce from \Eq{M_N} that the periodic switching only leads to the enhancement of certain fluctuations that are already present in the corresponding equilibrium Hamiltonians. The periodic time-dependence of the Hamiltonian leads to a selection of certain transitions that correspond to the absorption or emission of an odd number of energy quanta $\Omega$.

\section{Magnetization}

The dynamics of all quantities is determined by the Hamiltonian in \Eq{eff_Hamiltonian}. Due to the Emery-Kivelson transformation, the instantaneous value of the magnetization $P(t)$ of the impurity spin
\beq
	P(t)=\langle S_z(t) \rangle
\eeq
itself appears in the Hamiltonian. Therefore, the magnetization provides the full access to the time evolution in the present time-dependent setup. As it will turn out in the following, this does not imply that the problem has to be solved self-consistently. In contrast, the time evolution of the magnetization is independent of the potential scattering term appearing in the effective Hamiltonian as already mentioned below Eq.~(\ref{eff_Hamiltonian}). Due to \Eq{S_z_dd}, the magnetization is connected to the occupation $n_d(t)=\langle d^\dag(t) d(t) \rangle$ of the local $d$-level:
\beq
	P(t)=n_d(t)-\frac{1}{2}.
\eeq
As a consequence of the periodicity of the Hamiltonian, it is convenient to represent time coordinates in the following way:
\beq
	t=n\tau + s \quad ,n\in \mathbb{N}, s\in [0,\tau/2].
\eeq
For $s\in[\tau/2,\tau]$ the $S_z$ operator is constant due to the switch off of the spin dynamics. Therefore, the formulas in the following will always be presented for $s\in[0,\tau/2]$. Using the formulas in \Eq{M_N} and the fact that the matrices $\mathcal{M}^{(N)}$ are unitary, it can be shown that the magnetization decays exponentially in time:
\beq
	P(t)=P(0)\: e^{-n\Delta \tau} e^{-2 \Delta s},
\label{magnetization}
\eeq
where the time scale is set by the Kondo scale $t_K=1/T_K=1/(\pi w \Delta)$. For a single interaction quench, the magnetization $P_{\text{qu}}(t)$ equals:\cite{Leggett,Guinea,Lesage,Lob_dyn}
\beq
	P_{\text{qu}}(t)=P(0)\: e^{-2\Delta t}.
\label{magnetization_quench}
\eeq
Comparison with the result in \Eq{magnetization} reveals that the periodic driving affects the impurity spin orientation only by reducing the total time during which the spin dynamics in the Kondo Hamiltonian is switched on. The initial local spin polarization is transferred to the conduction band and flows away from the central region to infinity. Note that the magnetization is independent of the intermediate time evolution generated by the potential scattering Hamiltonian. 

In the limit of fast switching, $\tau \to 0$, the magnetization decays exponentially in time
\beq
	P_{\tau \to 0}(t)=P(0) \: e^{-\Delta t}.
\label{magnetization_tau0}
\eeq
The associated rate $\Delta$, however, is smaller compared to the single quench case where it is equal to $2\Delta$, see \Eq{magnetization_quench}. This is surprising, as one might expect that the additional energy provided to the system by the periodic driving may open additional phase space for relaxation processes. The decrease in the rate by one half occurs simply because the spin dynamics that are the only source of relaxation of the magnetization are switched on only during half of the time.

The exact result in \Eq{magnetization_tau0} for the magnetization, however, contrasts the dynamics one obtains by a naive implementation of the Trotter formula. Following a suggestion by Eisler and Peschel\cite{Eisler_Peschel} the dynamics of a periodically quenched system in the limit $\tau \to 0$ is identical to that of an effective equilibrium Hamiltonian that can be obtained by applying the Trotter formula\cite{Trotter} to the time evolution operator $U$ over one period:
\beq
	U=e^{-iH_1\tau/2}e^{-iH_2\tau/2}\approx e^{-i \frac{H_1+H_2}{2}\tau} + \mathcal{O}(\tau^2).
\label{eq_Trotter}
\eeq
As a result, the effective Hamiltonian equals the time-averaged one, that is a resonant-level model with a hopping amplitude $V/2$ plus a potential scattering term. The potential scattering term, however, does not affect the spin dynamics. Therefore, it will be omitted in the following. A resonant-level model with hopping amplitude $V/2$ generates a decay of the magnetization at a rate $\Delta/2$ as $\Delta\propto V^2$, compare \Eq{magnetization_quench}, in contrast to the exact value $\Delta$.

From a mathematical point of view, the Trotter formula fails as both Hamiltonians $H_1$ and $H_2$ have to be self-adjoint in a mathematical sense, that is a stronger requirement than Hermitian. For bounded operators (in fact, all realistic models are equipped with a high-energy cutoff), self-adjointness is guaranteed, in contrast to unbounded operators as considered in this work. In fact, regarding a resonant-level model with a nontrivial hopping element $V_k(t)=V(t)\: e^{-(k/k_c)^2}$ where $V(t)=V\theta[\sin(\Omega t)]$ and $k_c$ a high-energy cutoff, one indeed observes that in the limit $\tau \to 0$ the magnetization converges to the result obtained by applying the Trotter formula, see Eq.~(\ref{eq_Trotter}).

Physically speaking, the inapplicability of the Trotter formula is a consequence of the creation of high-energy excitations in the fast driving limit. The typical excitations generated by the periodic driving are of the order of $\Omega=2\pi/\tau$ corresponding to the absorption and emission of quanta of the driving frequency, compare Fig.~(\ref{pic_MkdN}). As $\Omega \to \infty$ for $\tau\to 0$, the typical excitations carry high energies. If the system is not provided with a mechanism, such as a finite bandwidth, that suppresses or cuts off these high-energy excitations, they will exist in the system even for $\tau\to 0$. Clearly, such excitations are not present in equilibrium systems leading to the conclusion that this fast driven system cannot described by an equilibrium Hamiltonian reflecting the mathematical statement above.

Summing up, the Trotter formula is not applicable in this model with a flat hybridization function. Despite the fact that realistic models actually exhibit a bounded spectrum, the results presented in this work describe correctly the dynamics in the limit $\Omega \gg T_K$, provided that the physical cutoff $D$ is still much bigger than $\Omega$.

\section{Spin-spin correlation function}

A dynamical quantity that carries more information about the local properties of the Kondo model is the spin-spin correlation function:
\beq
	\langle S_z(t) S_z(t') \rangle = C(t,t')-\frac{i}{2} \chi(t,t'),
\eeq
where $C(t,t')=\frac{1}{2}\langle \{ S_z(t),S_z(t') \} \rangle$ denotes the symmetrized part and $\chi(t,t')=i\theta(t-t')\langle [ S_z(t),S_z(t')] \rangle$ the response function for $t>t'$. After an infinite number of periods $\tau$, a steady state develops in which all real-time correlation functions are invariant under a discrete time shift $\tau$ in all their time arguments. For the spin-spin correlation function this implies
\beq
	\langle S_z(t+\tau) S_z(t'+\tau) \rangle =\langle S_z(t) S_z(t') \rangle . 
\label{periodicity_property}
\eeq
Therefore, the time coordinate $t'$ can be restricted to the interval $[0,\tau]$. The average $\langle \ldots \rangle$ in the steady state is to be understood as
\beq
	\langle S_z(t) S_z(t') \rangle = \lim_{N \to \infty} \langle \Psi_0(N\tau) | S_z(t) S_z(t') | \Psi_0(N\tau) \rangle,
\eeq
where $|\Psi_0\rangle$ denotes the initial state. In the steady state where $\langle S_z(t)\rangle=\langle S_z(t') \rangle=0$, the spin-spin correlation function equals
\beq
	\langle S_z(t) S_z(t') \rangle=\langle \hat n_d(t) \hat n_d(t') \rangle -\frac{1}{4}.
\eeq
Using the formulas in \Eq{M_N}, the spin-spin correlation function, that is a four-point function in terms of the fermionic operators of the effective Hamiltonian, can be related to a two-point function:
\beq
	\langle S_z(t) S_z(t') \rangle=\langle d^\dag(t) d(t') \rangle^2,
\eeq
where
\begin{widetext}
\begin{eqnarray}
	\langle d^\dag(t) d(t') \rangle =& & \frac{1}{\pi} \int_0^\infty d\omega \: \frac{e^{-i \omega \Delta \tau}}{1+\omega^2} \left[ 
	e^{-\Delta(s+t')}\frac{1-2\cos(\omega \Delta \tau/2)e^{-\Delta \tau/2}+e^{-\Delta \tau}}{1-2\cos(\omega \Delta \tau)e^{-\Delta \tau/2}+e^{-\Delta \tau}} \right.
	+\left(e^{-i\omega \Delta s}-e^{-\Delta s} \right)\left(e^{i\omega \Delta t'}-e^{-\Delta t'} \right)
	\nonumber \\
	& & \left. + e^{-\Delta s} \left( e^{-i \omega \Delta t'} - e^{-\Delta t'} \right) \frac{e^{-i\omega \Delta \tau/2}-e^{-\Delta \tau/2}}{e^{-i\omega \Delta \tau}-e^{-\Delta \tau/2}}
	+ e^{-\Delta t'} \left( e^{i \omega \Delta s} - e^{-\Delta s} \right) \frac{e^{i\omega \Delta \tau/2}-e^{-\Delta \tau/2}}{e^{i\omega \Delta \tau}-e^{-\Delta \tau/2}}
	\right].
\end{eqnarray}
\end{widetext}
Remarkably, the real part of the $\langle d^\dag (t) d(t') \rangle$ correlator can be calculated analytically
\beq
	\langle \left\{ d^\dag(t),d(t') \right\} \rangle=e^{-n \Delta \tau/2} e^{-\Delta(s-t')}
\eeq
matching precisely the result of Langreth and Nordlander\cite{Langreth} who derived a general formula of this correlator for an arbitrarily time-dependent resonant-level model. Inserting the time-dependence of the hopping amplitude $V(t)$ of the present setup into their result shows perfect agreement. As the anticommutator of two time-evolved fermionic single-particle operators is independent of the state if the Hamiltonian is quadratic, the full information about the influence of the steady state onto the spin-spin correlation function is contained in the imaginary part of the $\langle d^\dag(t) d(t') \rangle$ correlator. The imaginary part is not accessible analytically, but can be evaluated numerically. A representative plot of $C(t,t')$ and $\chi(t,t')$ is shown in Fig.~\ref{pic_spsp}. A detailed discussion will be given below.
\begin{figure}
\includegraphics[width=0.95\linewidth]{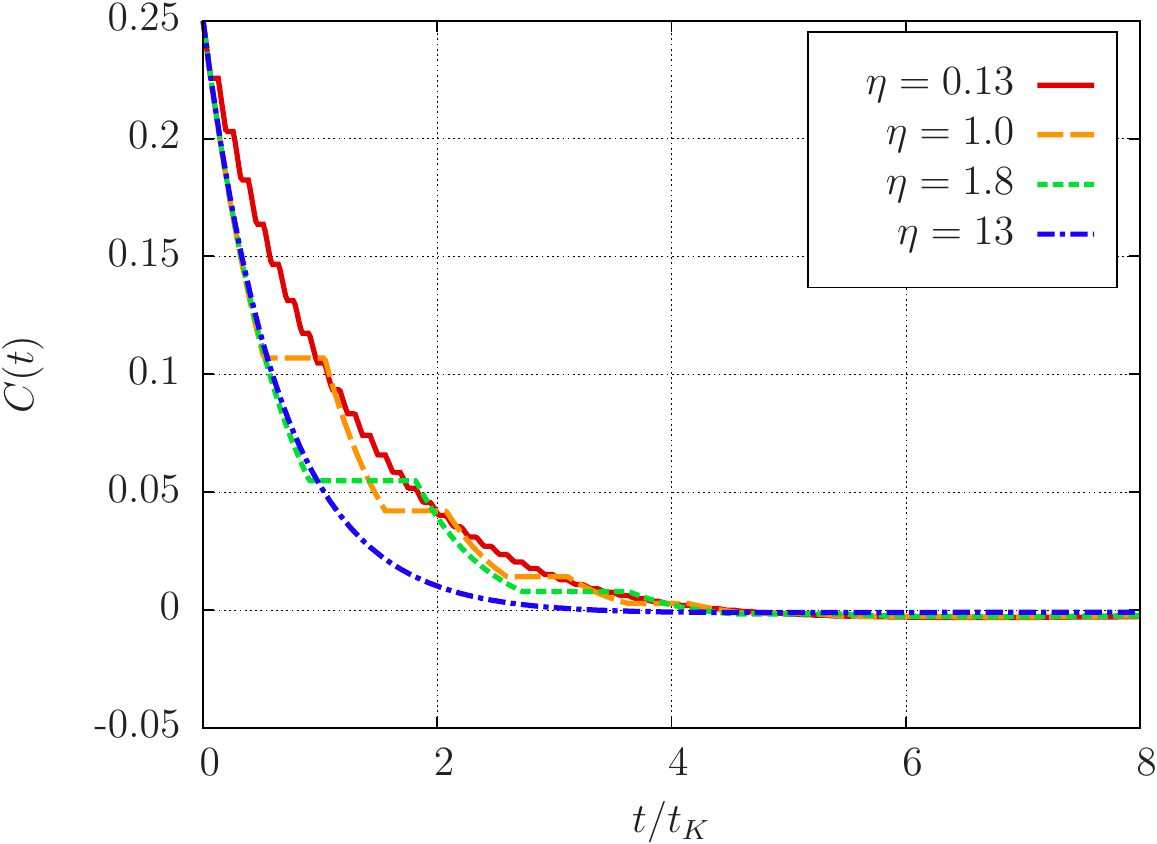}~\\~\\
\includegraphics[width=0.95\linewidth]{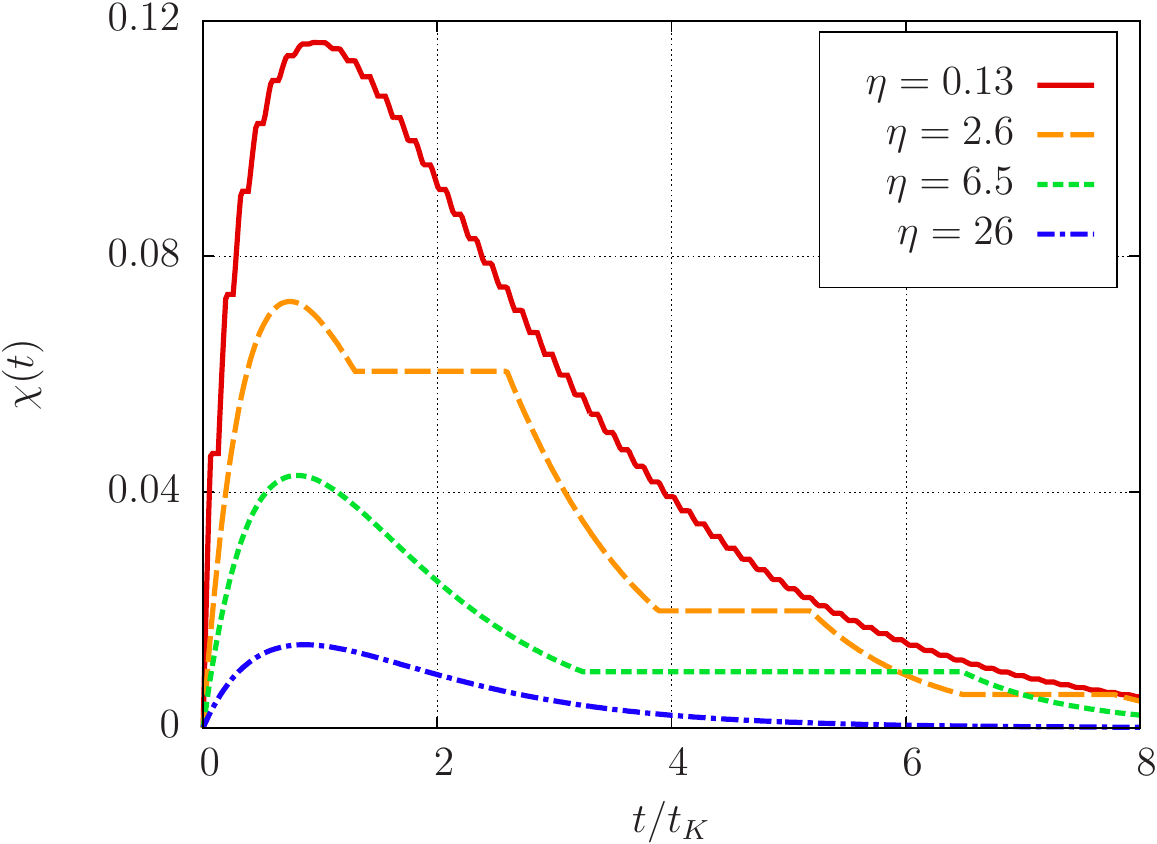}
\caption{Universal curves for the symmetrized correlation function $C(t)=C(t,0)$ and the response function $\chi(t)=\chi(t,0)$ for different values of the driving rate $\eta$ at zero waiting time $t'=0$. The parameter $\eta=\tau/t_K$ compares the speed of the external driving $\tau$ with the internal Kondo time scale $t_K=1/T_K$.}
\label{pic_spsp}
\end{figure}

Due to the appearance of a new scale in the present periodic driving setup, the universal description of the spin-spin correlation function gets modified in comparison to its equilibrium form:
\beq
	\langle S_z(t) S_z(t') \rangle=F\left( \frac{t}{t_K}, \frac{t'}{t_K},\frac{\tau}{t_K} \right),
\eeq
as one can expect by dimensional analysis. As a consequence, the Kondo scale, though an equilibrium quantity, remains the only relevant time and energy scale as it was shown for the conductance through a Kondo impurity in the work by Kaminski \textit{et al.}.\cite{Kam_prl,Kam_prb} The period $\tau$ only appears in combination with the Kondo time scale $t_K=1/T_K$ defining the ratio:
\beq
	\eta=\frac{\tau}{t_K}=\tau T_K.
\eeq
Therefore, it is only important how fast the system is driven compared to the internal time scale $t_K$.
~\\

\textit{The limit of long switching times:} In the limit of large periods $\tau$, the behavior of the local correlation functions is accessible by general arguments. Roughly speaking, the system is able to relax during each half period. Initially, the system is prepared in one of the ground states $|\Psi_0\rangle$ of the noninteracting Hamiltonian that are product states of the Fermi sea for the conduction-band electrons and a spin wave function for the local level. Switching on the interaction in the Kondo model creates local excitations in the vicinity of the local spin and the Kondo singlet forms. The excitations generated by the quench delocalize and flow away from the central region to infinity such that they cannot contribute to local properties any more. Therefore, the time evolved state looks like the ground state for local observables after a sufficiently long time. As was emphasized in Ref.~\cite{Lob_dyn}, however, the state $|\Psi_0\rangle$ can never develop into the true ground state of the Kondo model as the overlap of both wave functions is constant in time. Nevertheless, the time evolved state $|\Psi_0\rangle$ is essentially equivalent to the true ground state as far as expectation values of local observables such as the $S_z$ operator are concerned. All statements about relaxation of the state itself in the following are to be understood in this sense. After half of the period, the interaction in the Kondo model is switched off, thereby destroying the Kondo singlet. The excitations that are created by breaking up the Kondo singlet delocalize as argued before and the system evolves into the ground state of the noninteracting Hamiltonian, that is a product state of the Fermi sea and a spin wave function with zero magnetization. Therefore, the system at the moment of the second switch on of the interaction is prepared as initially up to a change in the local spin wave function. As a result, the system behaves as for a single interaction quench, a situation that has already been addressed by Lobaskin and Kehrein.\cite{Lob_dyn,Lob_eff} Analytically, the spin-spin correlation function transforms into:
\begin{eqnarray}
	\langle S_z(t) S_z(t') \rangle \stackrel{\tau \to \infty}{\longrightarrow} \left[ \frac{1}{2}e^{-\Delta(t-t')} -i \left( s(t-t') \right.\right. \nonumber \\
	\left.\left.-e^{-\Delta t'} s(t) + e^{-\Delta t} s(t') \right) \right]^2,
\label{spsp_large_tau}
\end{eqnarray}
where $s(t)=\pi^{-1}\int_0^\infty d\omega \: \sin(\omega \Delta t)/(\omega^2+1)$. This result matches precisely the result obtained by Lobaskin and Kehrein~\cite{Lob_dyn} for a single interaction quench.
~\\

\textit{The limit of fast switching:} In the opposite limit $\tau \to 0$, there exists no general argument capturing the dynamics in the resulting steady state. The sole reason for this is that the Trotter formula is not applicable in this model as explained in Sec.~IV. Otherwise, the dynamics in the steady state would be governed by an effective equilibrium Hamiltonian. Nevertheless, the fast driving generates dynamics similar to equilibrium as the system is not able to follow the fast externally prescribed perturbation.

Performing the limit $\tau \to 0$, the spin-spin correlation function reduces to:
\beq
	\langle S_z(t) S_z(t') \rangle\stackrel{\tau\to0}{\longrightarrow}\left[ \frac{1}{2}e^{-\Delta (t-t')/2}-i \frac{1}{2}s((t-t')/2) \right]^2.
\label{tau_zero}
\eeq
Thus, it only depends on the time difference signaling the similarity to an equilibrium problem in the sense that time-translational invariance is restored. The equilibrium spin-spin correlation function equals:\cite{Lob_dyn}
\beq
	\langle S_z(t) S_z(t')\rangle_{eq}= \left[ \frac{1}{2}e^{-\Delta (t-t')} - i s(t-t') \right]^2.
\label{spsp_correlator_equ}
\eeq
Comparing the result of Eq.~(\ref{tau_zero}) with the equilibrium case, one first observes that the time argument is scaled by a factor of $1/2$. As for the magnetization, this can be understood by the fact that the spin operators evolve nontrivially only during half of the time. Additionally, a prefactor of $1/2$ appears in front of the function $s$ such that the real and imaginary part of the $\langle d^\dag(t)d(t') \rangle$ correlator transform qualitatively different in the limit $\tau \to 0$. Thus, the fluctuation-dissipation theorem, a signature of equilibrium systems, is violated. This then leads to the conclusion that  it is impossible to find an equilibrium Hamiltonian generating the same dynamics.

Nevertheless, other quantities such as the dynamical spin susceptibility, as it will be shown in Sec.~VI,  can be related to equilibrium Hamiltonians. The inapplicability of the Trotter formula does not exclude the possibility to find equilibrium behavior, it only excludes the possibility to find a unique equilibrium Hamiltonian describing the correct dynamics for all observables. Concluding, the steady state in the limit $\tau \to 0$ provides an example for the emergence of new quantum states similar to equilibrium states but with new properties that are not accessible by equilibrium thermodynamics.
~\\

\textit{The asymptotic long-time behavior:} In equilibrium, the spin-spin correlation function exhibits a characteristic algebraic long-time behavior at zero temperature:
\beq
	\langle S_z(t) S_z(t') \rangle_{eq}\stackrel{t-t'\to \infty}{\longrightarrow}  -w^2\left[\frac{t_K}{t-t'} \right]^2 .
\eeq
At finite temperatures, the decay is exponential due to the smearing of the Fermi surface. Therefore, the long-time behavior of the spin-spin correlation function can serve as a measure whether the excitations that are created by the periodic driving are equivalent to those induced by a finite temperature. In this case, the steady state may be characterized by relating it to an equilibrium system at an effective temperature, a concept that has been widely used recently.\cite{Lob_eff,Nordl_Kondo_develop,Nordl_ac,Mitra}

In the present periodic time-dependent setup, the asymptotic long-time behavior of the spin-spin correlation function can be determined analytically:
\beq
		\langle S_z(t) S_z(t') \rangle \stackrel{t-t'\to \infty}{\longrightarrow}  -w^2\left[\frac{t_K}{t-t'}\right]^2.
\label{spsp_long_times}
\eeq
Surprisingly, the long-time behavior of the spin-spin correlation function is universal in the sense that it is completely independent of the external driving. Moreover, it precisely matches the equilibrium behavior at zero temperature. As the algebraic decay in equilibrium is caused by the sharp Fermi surface at zero temperature, one can conclude that the periodic driving is not able to smear the Fermi surface or to at least locally heat up the system excluding the concept of effective temperature. Furthermore, the low-energy excitations in the immediate vicinity of the Fermi level that are relevant for the long-time behavior are unaffected by the periodic driving. The excitation spectrum involves excitations of multiples of the driving frequency that emerge from processes where the local spin is flipped by simultaneously creating collective spin excitations in the fermionic reservoir with energies of odd multiples of the driving frequency $\Omega$, as can be seen in Fig.~\ref{pic_MkdN}. Therefore, the excitation spectrum is of discrete character, in contrast to the excitation spectrum induced by temperature.

\section{Dynamical spin susceptibility}

In the steady state, the magnetization of the impurity spin vanishes due to \Eq{magnetization}. A local spin polarization can be induced by applying a magnetic field to the local spin. In the linear response regime one obtains:
\beq
	\langle S_z(t) \rangle_h=\langle S_z(t) \rangle + \int_{-\infty}^\infty dt' \: \chi(t,t') h(t').
\eeq
Expectation values without an index $h$ are to be evaluated with respect to the unperturbed Hamiltonian. In the steady state, the magnetization vanishes such that the expectation value for the local spin polarization is solely determined by the response function
\beq
	\chi(t,t')=i\theta(t-t') \langle \left[ S_z(t), S_z(t') \right] \rangle.
\eeq
In equilibrium, the response function $\chi(t,t')$ only depends on the time difference thereby establishing a spectral representation of only one frequency argument whose imaginary part $\chi''(\epsilon)$ is called the dynamical spin susceptibility. It shows a peak located near the Kondo temperature $T_K$ that can be associated with the Kondo singlet. The existence of such a peak in a non-equilibrium setting can also be interpreted as a signature for the presence of the Kondo effect as a whole, as the Kondo singlet is just one manifestation of this coherent many-body phenomenon. Moreover, in equilibrium, the fluctuation-dissipation theorem holds that connects the dynamical spin susceptibility with the local spin fluctuation spectrum. In systems out of equilibrium, the Fluctuation-Dissipation theorem is violated, as can be seen explicitly for a single interaction quench scenario in the Kondo model.\cite{Lob_eff}

For periodic time-dependent Hamiltonians there also exists a preferable spectral decomposition. Due to the periodicity property  in \Eq{periodicity_property}, it is suitable to define two new time arguments:
\beq
	\tave=\frac{t+t'}{2}, \: \trel=t-t'.
\eeq
Expressing $\chi(t,t')$ in these coordinates, $\chi(\tave,\trel)$ is invariant under the transformation $\tave \to \tave +\tau$ such that a Fourier series expansion in the coordinate $\tave$ can be performed. Therefore, one can spectrally decompose $\chi$ in the following way that is usually referred to as the Wigner representation:
\begin{eqnarray}
	& & \chi_{n}(\varepsilon)= \nonumber\\
	& & =\frac{1}{\tau}\int_0^\tau d\tave e^{in\Omega \tave} \int d\trel  e^{i \varepsilon \trel} \chi\left(\tave +\frac{\trel}{2},\tave-\frac{\trel}{2}\right) \nonumber \\
	& & =\frac{1}{\tau}\int_0^\tau d\tave e^{in\Omega \tave} \int d\trel  e^{i \overline{\varepsilon} \trel} \chi\left(\tave +\trel,\tave\right) \nonumber \\
	& & =\frac{1}{\tau} \int_0^\tau d\tave e^{in\Omega \tave} \chi(\tave,\overline{\varepsilon})
\label{Wigner_representation}
\end{eqnarray}
where $\overline{\varepsilon}=\varepsilon+n\Omega/2$. With each component $n$, one can associate the behavior of the quantity $\chi$ due to processes where $n$ quanta of the driving frequency $\Omega$ are absorbed ($n>0$) or emitted ($n<0$). The quantity $\chi(\tave,\varepsilon)$ can be interpreted as the spectral decomposition of $\chi$ at a given point $\tave$ in time. The $n=0$ component of $\chi_{n}$ simply is the time average of the quantity $\chi(\tave,\varepsilon)$.

Applying a small sinusoidal magnetic field $h(t)=h_0 \sin(\Omega_0 t)$ to the local spin in the steady state, linear response theory predicts for the magnetization of the impurity spin $\langle S_z(t) \rangle_h$ in the presence of the small perturbation $h$:
\begin{eqnarray}
	\langle S_z(t) \rangle_h=h_0\sum_{n}\left[ \chi'_n(\Omega_0+n\Omega/2) \sin[(\Omega_0+n\Omega)t] \right. \nonumber \\
	\left.-\chi''_n(\Omega_0+n\Omega/2) \cos[(\Omega_0+n\Omega)t] \right].
\end{eqnarray}
Remarkably, this expression contains a static contribution if $\Omega_0=m\Omega$, $m\in\mathbb{Z}\backslash\{0\}$. Therefore, it is possible to align the local spin on average by applying a sinusoidal magnetic field that itself contains no static contribution:
\beq
	\overline{\langle S_z(t) \rangle_h}=\chi''_m(m\Omega/2).
\eeq
Here, $\overline{\langle \dots \rangle}$ denotes the time average. As the spin dynamics are switched off during the second half period, the external magnetic field is not able to influence the local spin magnetization in this time window regardless of its time dependence. Therefore, the magnetic field must be capable of polarizing the local spin during the first half period in order to induce a static component. This can be achieved by $h(t)=h_0\sin(\Omega_0 t)$ with $\Omega_0=m\Omega$. 

Note that a definite order of taking limits is implicitly prescribed in this linear response calculation. First, the system is evolved into the steady state that is established after an infinite number of periods. Afterwards, an additional sinusoidal magnetic field is applied that acts for an eventually infinite amount of time. Therefore, linear response theory only provides the information of how an additional infinitesimal magnetic field influences the local spin properties after the steady state has already been established. It does not necessarily describe the properties of a system whose interaction is periodically switched on and off in presence of a magnetic field, as the two involved limiting processes may not commute.
~\\

\begin{figure}
\includegraphics[width=0.95\linewidth]{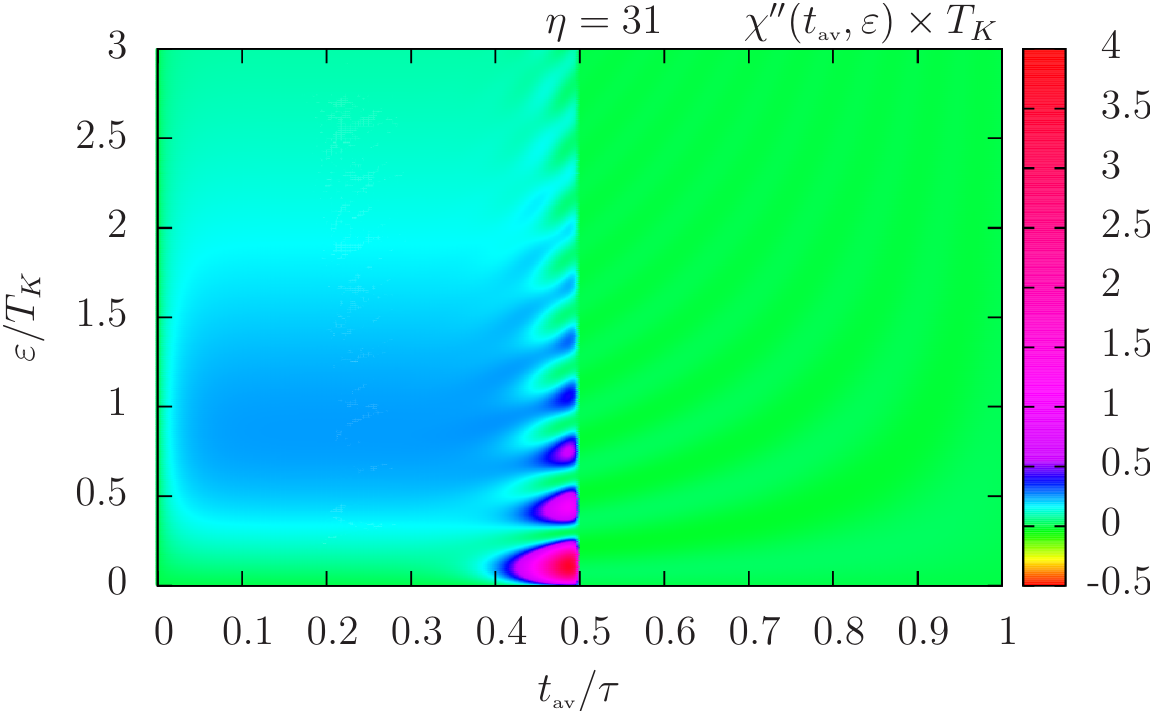}~\\~\\
\includegraphics[width=0.95\linewidth]{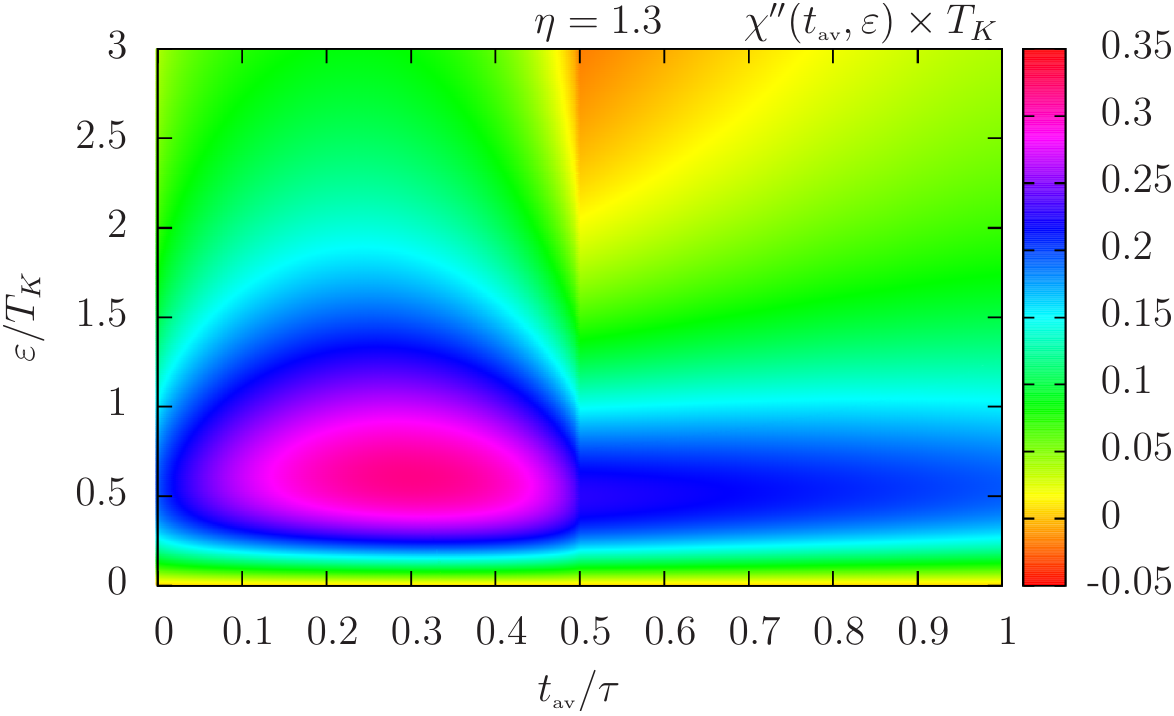}~\\~\\
\includegraphics[width=0.95\linewidth]{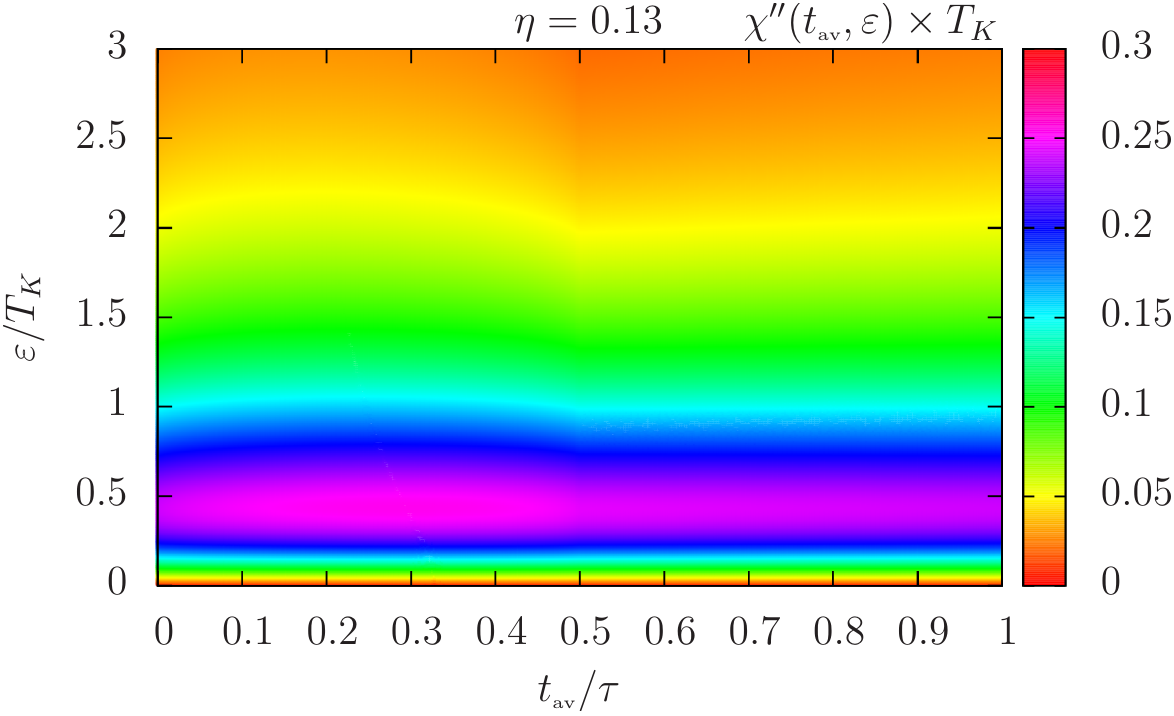}
\caption{False color plots of the dynamical susceptibility $\chi''(\tave,\varepsilon)$ at a given point $\tave$ in time for different values of the parameter $\eta$. The upper plot shows the behavior for slow driving consistent with a single interaction quench picture. The lower plot displays the fast driving behavior that is similar to an equilibrium problem. The plot in the middle shows the intermediate regime of competing time scales.}
\label{pic_dynsusz}
\end{figure}

\textit{Results for $\chi''(\tave,\varepsilon)$: }In Fig.~\ref{pic_dynsusz}, results for the dynamical spin susceptibility $\chi''(\tave,\varepsilon)$ are shown for different values of the driving frequency $\Omega$. The first plot displays the case of slow driving, i.e., $\tau \gg t_K$, and reveals again the single quench dynamics as argued before. After a transient regime on a time scale $t_K$, that is the time scale for the buildup of the Kondo effect\cite{Nordl_Kondo_develop}, the dynamical spin susceptibility approaches its equilibrium shape with a peak located at $\varepsilon\approx T_K$ representing the buildup of the Kondo singlet and therefore of the Kondo effect. Near the half period boundary, however, a new structure emerges. At $\tave\approx \tau/2-t_K$, a whole sequence of strong peaks appears that cannot be explained by the simple single interaction quench picture as this coherent phenomenon is solely caused by the periodic driving. Approaching $\tave=\tau$ these peaks rapidly decrease and disappear. Note, that the dynamical spin susceptibility is not discontinuous at $\tave=\tau/2$ as it might seem. The collapse near $\tau/2$, however, is so fast that it cannot be resolved by the pictures presented here.

In the case of fast switching, the lower plot in Fig.~\ref{pic_dynsusz}, the dynamical spin susceptibility stays nearly constant over the whole period. The system is not able to adapt to the fast external perturbation, as it is faster than the internal time scale $t_K$, on which the system is able to react. Remarkably, the shape of the spectral decomposition at any time point $\tave$ resembles the shape of an equilibrium dynamical spin susceptibility with a reduced Kondo temperature $T_K/2$. It was shown in \Eq{tau_zero} that the response function $\chi_{\tau \to 0}(t, T_K)$ in the limit $\tau \to 0$ is identical to half of an equilibrium response function $\chi_{eq}(t/2,T_K)/2$ at a rescaled time argument $t/2$ and identical Kondo temperature $T_K$. Here, the Kondo temperature has been included explicitly for convenience. Such a relation for functions in time implies the following relation for the corresponding Fourier transforms:
\beq
	\chi_{\tau \to 0}(\varepsilon,T_K)=\chi_{eq}(2\varepsilon,T_K).
\eeq
For the dynamical spin susceptibility, this statement can be rewritten in terms of a modified Kondo temperature:
\beq
	\chi''_{\tau \to 0}(\varepsilon,T_K)=2\chi''_{eq}(\varepsilon,T_K/2).
\eeq
The equilibrium dynamical susceptibility at zero temperature is known exactly, such that\cite{Guinea_dynsusz,Leggett}
\begin{eqnarray}
	\chi''_{\tau \to 0}(\varepsilon)=\frac{\Delta^2}{2\pi}\frac{1}{\varepsilon^2+\Delta^2}\left[ \frac{1}{2\varepsilon}\ln\left(1+\left(\frac{2\varepsilon}{\Delta} \right)^2\right) \right. \nonumber\\
	\left.+\frac{1}{\Delta}\arctan\left(\frac{2\varepsilon}{\Delta}\right) \right].
\end{eqnarray}
A Kondo singlet is present even for fast periodic driving with a reduced binding energy $T_K/2$. Therefore, one can speculate that the Kondo effect itself survives with a rescaled Kondo temperature $T_K/2$.~\cite{TK4} Although this analogy suggests the equivalence to an effective equilibrium problem, this is not valid in a strict sense, since it was shown, compare \Eq{tau_zero}, that one cannot find an effective equilibrium Hamiltonian that generates the same dynamics. Nevertheless, it is remarkable that the dynamical spin susceptibility, that is a measurable quantity in principle, shows an equilibrium-like behavior of a Kondo model with reduced Kondo temperature in this limit.

The robustness of the Kondo effect even for $\Omega \gg T_K$ has also been observed by Nordlander and coworkers\cite{Nordl_ac} where the time-averaged local spectral density shows a Kondo resonance in the limit $\Omega \gg T_K$.

The plot in the middle of Fig.~\ref{pic_dynsusz} shows the intermediate regime where both the external time scale $\tau$ and the internal time scale $t_K$ are of the same order. In this case, the system is able to adapt partially to the externally prescribed perturbation. For small times $\tave$, one observes that the Kondo singlet tries to form, the dynamical spin susceptibility tends to approach its equilibrium profile. Near the half period boundary where the interaction is switched off, however, the dynamical spin susceptibility collapses onto a curve with a peak located approximately at $T_K/2$ as in the fast driving case.
~\\

\begin{figure}
\includegraphics[width=0.95\linewidth]{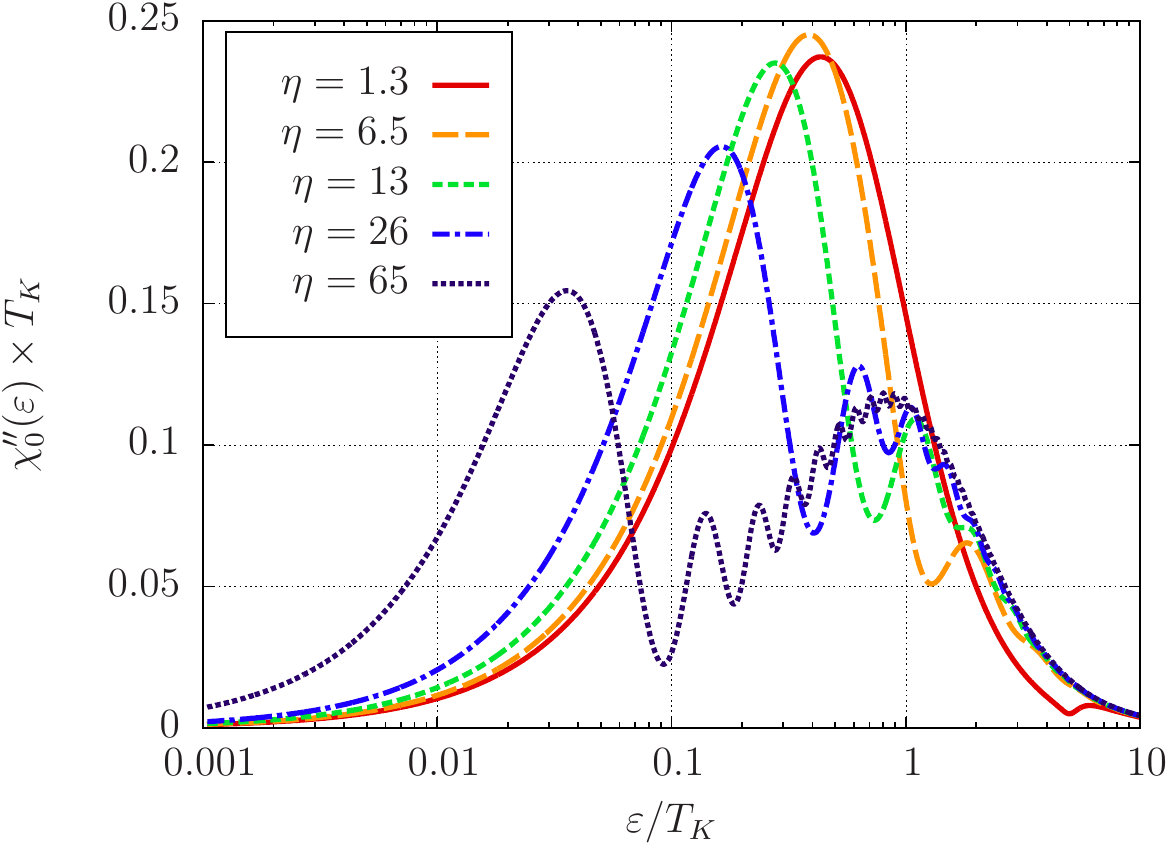}~\\~\\
\includegraphics[width=0.95\linewidth]{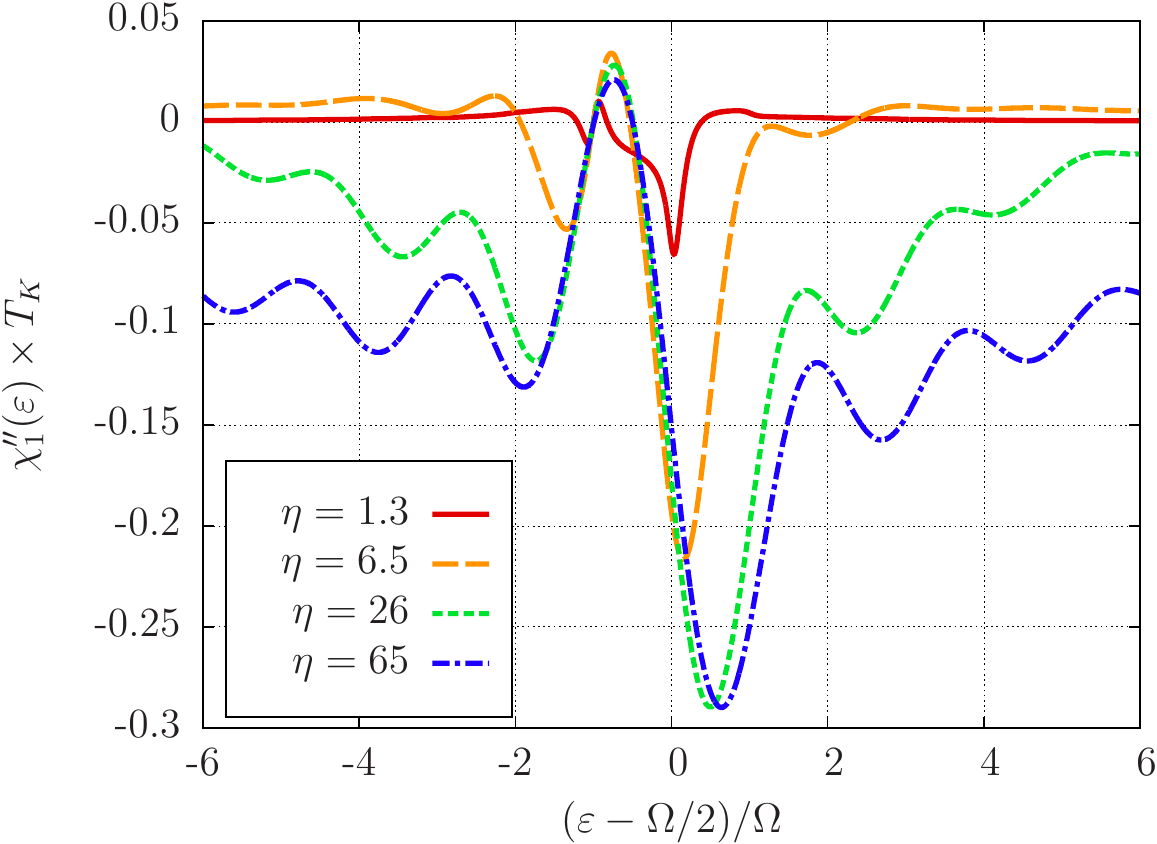}~\\~\\
\includegraphics[width=0.95\linewidth]{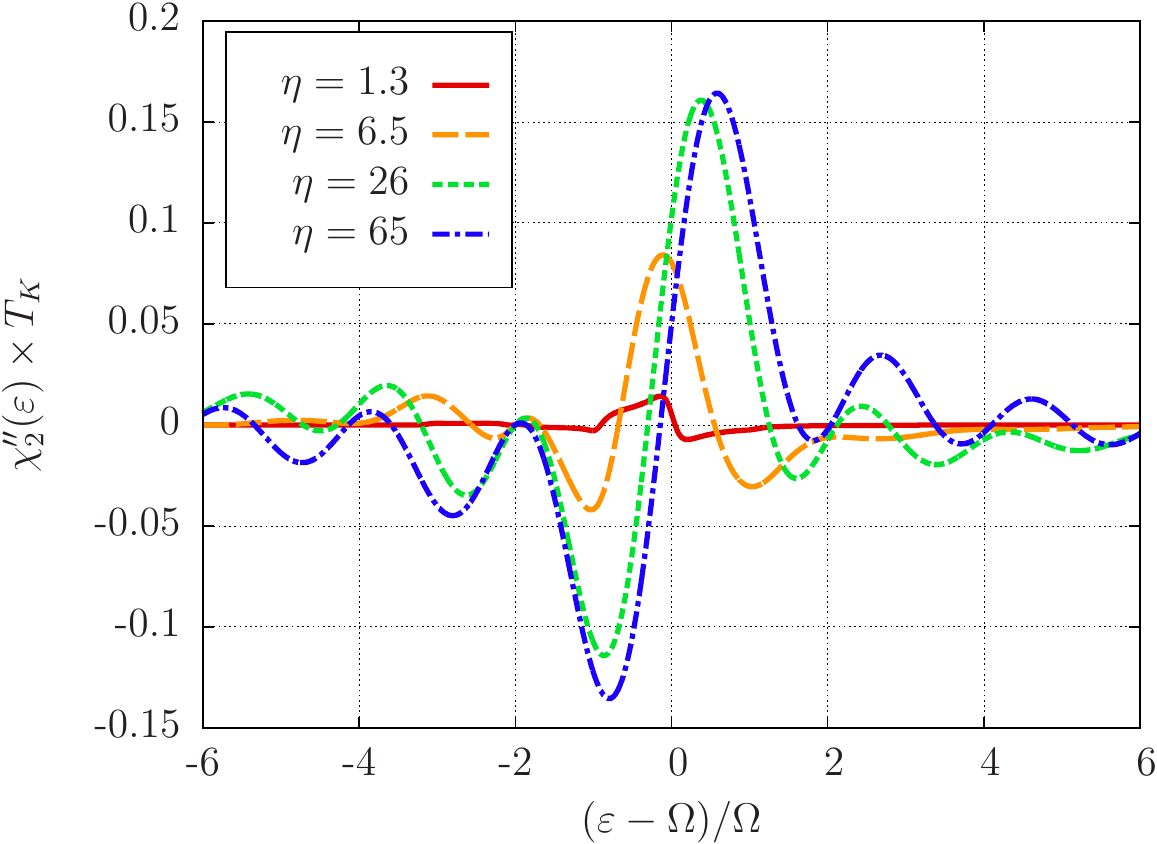}~\\~\\
\caption{Universal curves of the zeroth, first and second mode $\chi''_n(\varepsilon-n\Omega/2)$ for different values of the parameter $\eta$.}
\label{pic_Floquet_modes}
\end{figure}

\textit{Results for $\chi''_n(\varepsilon)$: }In Fig.~\ref{pic_Floquet_modes}, universal curves for the modes $\chi''_n(\varepsilon)$ are shown. In the upper plot of the $n=0$ component, one can nicely see the crossover from the fast to the slow driving regime. For fast driving, small $\eta$, one observes a peak located at $\varepsilon \approx T_K/2$, whose height and position decreases by increasing $\eta$. This signals a decrease in the binding energy as well as the stability of the associated Kondo singlet. For large $\eta$, this peak vanishes according to the single quench dynamics for $\tau=\infty$ where only a Kondo singlet of energy $T_K$ exists. Indeed, for intermediate $\eta$, a second peak emerges at $T_K$ with a height of only half of its equilibrium value, as half of the time the interaction is switched off. Thus, the $n=0$ component of the dynamical susceptibility shows the crossover from a Kondo singlet of binding energy $T_K$ at $\tau=\infty$ to a Kondo singlet with binding energy $T_K/2$ for $\tau=0$. The oscillations appearing for large $\eta$ originate in the periodic structure near the half period boundary of $\chi(\tave,\varepsilon)$ that can be seen in Fig.~\ref{pic_dynsusz}.

The $n=1$ and $n=2$ modes that correspond to processes where one or two energy quanta $\Omega$ are absorbed are shown in the lower two plots of Fig.~\ref{pic_Floquet_modes}. Due to the relation
\beq
	\chi''_n(\varepsilon)=-\chi''_{-n}(-\varepsilon),
\eeq
these results can be extended straightforwardly to the $n=-1,-2$ modes. 

For small periods $\tau$, all modes $n\not=0$ vanish due to the equilibrium-like dynamics where absorption and emission of photons is not taking place. As $\Omega\to\infty$ for $\tau\to 0$, the absorption or emission of an energy quantum $\Omega$ involves high energy transfers that are cut off by the finite width of the local hybridized $d$ level. The equilibrium spin fluctuations of the conduction-band electrons in the Kondo model at large energies are suppressed due to energy conservation. As mentioned in Sec. III, the only influence of the periodic driving is the enhancement of certain fluctuations that are present in the equilibrium versions of the involved Hamiltonians. If the equilibrium fluctuations already vanish, their enhancement due to the periodic driving is negligible, too.

The larger the parameter $\eta$, the larger becomes the response for processes under absorption and emission, as the excitations generated by the periodic driving now turn out to be excitations of lower energy as $\Omega$ decreases. In the limit $\tau\to \infty$, however, the $n\not=0$ components are expected to vanish as in the $\tau\to 0$ limit, since the system is able to relax during each half period and locally behaves as in equilibrium nearly all of the time. The behavior for the $n=1$ and $n=2$ components is qualitatively different for energies above or below $n\Omega/2$. The absorption or emission of one photon is likely for the $n=1$ mode whereas the $n=2$ mode decays to zero much faster away from its maximum. The different properties of the even and odd modes can be traced back to \Eq{odd_freq}, stating that the square wave driving can be thought of as a non-monochromatic driving including all frequencies of odd multiples of $\Omega$. 

\section{Conclusions}

In this work, the steady state in a periodically driven Kondo model in the Toulouse limit has been investigated by analyzing exact analytical results for the local spin dynamics at zero temperature. Remarkably, the analysis revealed a universal asymptotic long-time behavior of the spin-spin correlation function, compare Eq.~(\ref{spsp_long_times}), independent of the driving matching precisely the zero temperature equilibrium result. This universality originates in the discrete character of the excitation spectrum that leaves the low-energy excitations in the immediate vicinity of the Fermi level unchanged.

For large switching times, $\tau/t_K \to \infty$ where $t_K$ is the equilibrium Kondo time scale, the local observables behave as for a single interaction quench, since the system is able to relax during each half period. In the opposite limit, $\tau/t_K \to 0$, the system is not able to follow the fast externally prescribed perturbation. Although the dynamics is ``equilibrium-like'' (in the sense that time-translational invariance is restored, e.g. Eq.~(\ref{tau_zero})), it is impossible to find an effective equilibrium Hamiltonian generating the same dynamics. This can be seen in the discussion below Eq.~(\ref{spsp_correlator_equ}) were it was shown that the fluctuation-dissipation theorem is violated for the local spin observable. Consequently, the steady state created by the periodic time-dependent setup provides an example for the possibility to reach new quantum states in systems out of equilibrium  that are not accessible by equilibrium thermodynamics. The reason for the existence of this unusual steady state is the inapplicability of the Trotter formula in the present periodic time-dependent setup. This observation might also be useful for other periodically driven systems.

Remarkably, in the Toulouse limit the Kondo effect is robust against the periodic driving as long as the driving frequency is unable to induce charge fluctuations in the quantum dot. Even in the case of fast driving $\Omega \gg T_K$ the Kondo singlet survives. The associated binding energy $T_K/2$ is reduced to half of its equilibrium value. For intermediate $\tau$, a crossover is observed from a Kondo singlet of binding energy $T_K/2$ in the fast switching limit to a Kondo singlet of binding energy $T_K$ for long switching times. 

It would be interesting to investigate related questions for
the Kondo model away from the Toulouse limit, where one also expects that the flow to the strong-coupling fixed point itself is affected
by the periodic switching. Likewise it would be important to understand whether other periodically driven quantum impurity models
show similar dynamics. Work along these lines is in progress. The exact analytical results derived in this work will be important
benchmarks for these investigations.

\begin{acknowledgments}
This work was supported through SFB~484 of the Deutsche 
Forschungsgemeinschaft (DFG), the Center for Nanoscience (CeNS)
Munich, and the German Excellence Initiative via the
Nanosystems Initiative Munich (NIM).
\end{acknowledgments}

\appendix
\section{}
\label{App_A}

In this appendix, the procedure to obtain the solution to the matrix multiplication problem is presented. For that purpose, the Green's functions for the resonant-level model $\mathcal{G}$ and the potential scattering Hamiltonian $\mathcal{G}^{(n)}$ have to be derived, for example, by using the equations of motion approach. The matrices $\mathcal{M}(n)=\mathcal{G}\mathcal{G}^{(n)}$ obey the following formulas:
\begin{eqnarray}
	\mathcal{M}_{dd}(n) & & =e^{-\Delta \tau/2}, \nonumber\\
	\mathcal{M}_{dk}(n) & & =\frac{V}{k-i\Delta}\left[ e^{-ik\tau/2} -e^{-\Delta \tau/2}\right], \nonumber \\
	\mathcal{M}_{kd}(n) & & =\frac{Ve^{-ik\tau/2}}{k-i\Delta}\left[ e^{-ik\tau/2} -e^{-\Delta \tau/2}\right], \nonumber \\
	\mathcal{M}_{kk'}(n) & & =\delta_{kk'}e^{-ik\tau} + \mathcal{L}_{kk'}+\mathcal{K}_{kk'}(n)
\end{eqnarray}
where the matrices $\mathcal{L}$ and $\mathcal{K}(n)$ are determined by:
\begin{eqnarray}
	\mathcal{L}_{kk'} & & =V^2 e^{-ik\tau/2} \left[ \frac{e^{-ik\tau/2}}{(k-k')(k-i\Delta)} \right. \nonumber \\
		& & +\left. \frac{e^{-\Delta \tau/2}}{(k-i\Delta)(k'-i\Delta)}+\frac{e^{-ik'\tau/2}}{(k'-k)(k'-i\Delta)} \right] \nonumber \\
	\mathcal{K}_{kk'}(n) & & =\frac{g\langle S_z (n\tau)\rangle e^{-ik\tau/2}}{1+ig\langle S_z (n\tau)\rangle L/2}\frac{e^{-ik\tau/2}-e^{-ik'\tau/2}}{k-k'}
\end{eqnarray}
For simplicity, the method to obtain $\mathcal{M}^{(N)}$ will be displayed by calculating the matrix element $\mathcal{M}^{(N)}_{dd}$. Using its definition in \Eq{def_matrix_M_N} one can write:
\begin{widetext}
\begin{eqnarray}
	\mathcal{M}^{(N)}_{dd} & = & \sum_{l_1,\ldots,l_{N-1}} \mathcal{M}_{dl_1}(1) \dots \mathcal{M}_{l_{N-1}d}(N)=\mathcal{M}^{(N-1)}_{dd} \mathcal{M}_{dd} +\sum_{l_1,\ldots,l_{N-3},k_{N-1}} \mathcal{M}_{dl_1}(1)\dots \mathcal{M}_{l_{N-3}d}(N-2) \mathcal{M}_{dk_{N-2}} \mathcal{M}_{k_{N-2}d}\nonumber \\
	& & + \sum_{l_1,\ldots,l_{N-3},k_{N-2},k_{N-1}}\mathcal{M}_{dl_1}(1) \dots \mathcal{M}_{l_{N-3}k_{N-2}}(N-2)\mathcal{M}_{k_{N-2}k_{N-1}}(N-1)\mathcal{M}_{k_{N-1}d}.
\label{M_N_dd_prel}
\end{eqnarray}
\end{widetext}
To proceed further, the following relations are required:
\begin{eqnarray}
	& & \sum_{k'} \mathcal{M}_{kk'}(n) \mathcal{M}_{k'd}e^{-i\lambda k'\tau}=e^{-i(\lambda+1)k\tau}\mathcal{M}_{kd}, \nonumber \\
	& & \sum_{k} \mathcal{M}_{dk}\mathcal{M}_{kd}=0
\end{eqnarray}
that originate in elementary properties of the matrix $\mathcal{M}$, namely that $\mathcal{M}$ as a function of $k$ exhibits no poles and that the exponentials $e^{-ik\tau/2}$ always allow to deform integration contours into the lower half plane. Using these formulas, one can directly show that the two sums appearing in \Eq{M_N_dd_prel} vanish. The resulting recursion formula is solved by:
\beq
	\mathcal{M}_{dd}^{(N)}=e^{-N\Delta \tau/2}.
\eeq
Analogously, the other matrix elements of $\mathcal{M}^{(N)}$ can be derived, although the evaluation of $\mathcal{M}_{kk'}^{(N)}$ is quite involved. Remarkably, all matrix elements with an index $d$ are independent of the potential scatterer. Only $\mathcal{M}^{(N)}_{kk'}$ includes a term that can be traced back to the potential scattering:
\beq
	\mathcal{K}_{kk'}^{(N)}=g\langle S_z(0)\rangle \sum_{n=0}^{N-1} \frac{e^{-ink\tau}e^{-i(N-n)k'\tau}}{e^{n\Delta \tau}+ig\langle S_z(0) \rangle L/2}.
\eeq
Here, the result for the magnetization, see \Eq{magnetization}, has been used.


\begin{thebibliography}{30}
\bibitem{TK4} Note that the time averaged effective Hamiltonian would generate a Kondo model with a Kondo temperature $T_K/4$ in contrast to the exact value $T_K/2$.
\end{thebibliography}
\end{document}